\documentstyle[aps]{revtex}

\begin{document}
\title{Spin structure functions and intrinsic motion of the constituents
\protect\footnote{%
to be published in Phys. Rev. D}}
\author{Petr Z\'avada}
\address{Institute of Physics, Academy of Sciences of the Czech Republic,\\
Na Slovance 2, CZ-182 21 Prague 8, \\
e-mail: zavada@fzu.cz \\
}
\date{January 4, 2002}
\maketitle

\begin{abstract}
The spin structure functions of the system of quasifree fermions on mass
shell are studied in a consistently covariant approach. Comparison with the
basic formulas following from the quark-parton model reveals the importance
of the fermion motion inside the target for the correct evaluation of the
spin structure functions. In particular it is shown, that regarding the
moment $\Gamma_{1}$, both the approaches are equivalent for the static
fermions, but differ by the factor $1/3$ in the limit of massles fermions ($%
m \ll p_0$, in target rest frame). Some other sum rules are discussed as
well.
\end{abstract}

\section{Introduction}

\label{sec1}Measuring of the nucleon spin structure functions represents an
important tool not only for better understanding of the nucleon internal
structure in the language of the QCD, but also for better understanding of
QCD itself. These functions contain an information, which is a crucial
complement to the structure functions obtained in the unpolarized deep
inelastic scattering (DIS) experiments.

The polarized experiments are more complex and difficult than the
unpolarized ones, nevertheless the last decade has brought remarkable
results also for the nucleon spin functions from the experiments at CERN
(EMC, SMC) and SLAC (E142, E143, E154, E155). And the new experiments are
running (HERMES) or are being under preparation (COMPASS). The data on
polarized $pp$ collisions are expected from the collider RHIC. For the
present status of the research in structure functions see e.g. \cite{dis},
the overview \cite{voss} and citation therein. The more formal aspects of
the polarized DIS are explained in \cite{efrem1}.

Also the interpretation and understanding of polarized structure functions
seem be more difficult. For an example, until now it is not well understood,
why the integral of the proton spin structure function $g_{1}$ is
substantially less, than expected from very natural assumption, that the
nucleon spin is generated by the valence quarks. Presently, there is a
tendency to explain the missing part of the nucleon spin as a contribution
of the gluons. It has been also suggested, that the quark orbital momentum
can play some role as well \cite{bro}-\cite{rit}.

The spin in general is a very delicate quantity, which requires
correspondingly precise treatment. It has been argued, that for correct
evaluation the quark contribution to the nucleon spin it is necessary to
take properly into account the intrinsic quark motion \cite{bro} - \cite%
{zav3}. Necessity of the covariant formulation of the quark - parton model
(QPM) for the spin functions has been pointed out in \cite{cfs}. These
requirements are not satisfied in the standard formulation of the QPM, which
is currently used for analysis and interpretation of the experimental data.

In this paper we shall attempt to demonstrate the role of the intrinsic
motion for the spin structure functions, using very simple model of the
system quasifree fermions on mass shell. The basic requirement is
consistently covariant formulation of the task for the system of fermions,
which are not static, being characterized by some momenta distribution in
the frame of their centre of mass. The spin structure functions of such
system are obtained in Sec. \ref{sec2} and the sum rules following from
these functions are shown in Sec. \ref{sec3}. In the Sec. \ref{sec4} a
comparison with the formulas of the standard QPM is done. The last section
is devoted to the short summary.

\section{Spin structure functions in covariant approach}

\label{sec2}Let us imagine a system of three quasifree charged fermions with
the spin $1/2$ and mass $m$, for which the following conditions are
satisfied:

1) The distribution of fermion momenta in the frame of their centre of mass
is described by some spherically symmetric function $G,$%
\begin{equation}
\int G(p_{0})d^{3}p=3;\qquad p_{0}=\sqrt{m^{2}+{\bf p}^{2}}.  \label{cr1}
\end{equation}%
The free fermion states are described by the spinors%
\begin{equation}
\psi _{p,\lambda }(x)=\frac{1}{\sqrt{\Omega }}u\left( p,\lambda \right) \exp
(-ipx);\qquad \int_{\Omega }\psi _{p,\lambda }^{\dagger }(x)\psi _{p,\lambda
}(x)d^{3}x=1,  \label{cra1}
\end{equation}%
where $\Omega $ is the normalization volume and%
\begin{equation}
u\left( p,\lambda \right) =\frac{1}{\sqrt{N}}\left( 
\begin{array}{c}
\phi _{\lambda } \\ 
\frac{{\bf p}{\bf \sigma }}{p_{0}+m}\phi _{\lambda }%
\end{array}%
\right) ;\qquad N=\frac{2p_{0}}{p_{0}+m},\qquad \phi _{\lambda }^{\dagger
}\phi _{\lambda }=1.  \label{crc1}
\end{equation}%
We assume%
\begin{equation}
\frac{1}{2}{\bf n\sigma }\phi _{\lambda }=\lambda \phi _{\lambda },\qquad
\lambda =\pm \frac{1}{2},  \label{crb1}
\end{equation}%
which means, that the spin projection of the fermion in its rest frame is $%
\pm 1/2$ in the given direction ${\bf n};\left| {\bf n}\right| =1.$

2) By $G_{\pm 1/2}\ $we denote function, which measures probability, that
fermion is in the state $\psi _{p,\pm 1/2}$, so that%
\begin{equation}
G(p_{0})=G_{+1/2}(p_{0})+G_{-1/2}(p_{0})  \label{cr2}
\end{equation}%
and we assume

\begin{equation}
\int \Delta G(p_{0})d^{3}p=1;\qquad \Delta G(p_{0})\equiv
G_{+1/2}(p_{0})-G_{-1/2}(p_{0}).  \label{crg3}
\end{equation}%
The difference $\Delta G$ consists of the corresponding contributions $%
\Delta h_{j}$ from the three fermions:%
\begin{equation}
\Delta G(p_{0})=\sum_{k=1}^{3}\Delta h_{k}(p_{0});\qquad \Delta
h_{k}(p_{0})\equiv h_{k,+1/2}(p_{0})-h_{k,-1/2}(p_{0}).  \label{crd3}
\end{equation}%
Later on, we shall need also the distribution%
\begin{equation}
H(p_{0})\equiv \sum_{k=1}^{3}e_{k}^{2}\Delta h_{k}(p_{0}),  \label{cre3}
\end{equation}%
where $e_{k}$ are the fermion charges.\qquad \qquad

What is the resulting spin (total angular momentum) related to the whole
system? Let us calculate the integral of the matrix elements%
\begin{equation}
\left\langle {\bf nj}\right\rangle =\int \int_{\Omega }\sum_{\lambda
}G_{\lambda }(p_{0})\left( \psi _{p,\lambda }^{\dagger }(x){\bf nj}\psi
_{p,\lambda }(x)\right) d^{3}xd^{3}p,\qquad  \label{cra3}
\end{equation}%
where the angular momentum ${\bf j}$ consists of the spin and orbital part 
\begin{equation}
j_{k}=\Sigma _{k}+l_{k}=\frac{1}{2}\left( 
\begin{array}{cc}
\sigma _{k} & 0 \\ 
0 & \sigma _{k}%
\end{array}%
\right) -i\varepsilon _{klm}p_{l}\frac{\partial }{\partial p_{m}}.
\label{crh3}
\end{equation}%
Since the total angular momentum ${\bf j}$ is a conserving quantity, which
commutes with the term ${\bf p}{\bf \sigma }$, a simple calculation gives%
\begin{equation}
\psi _{p,\lambda }^{\dagger }(x){\bf nj}\psi _{p,\lambda }(x)=\frac{1}{%
\Omega }\left( \lambda +\varepsilon _{klm}n_{k}p_{l}x_{m}\right) .
\label{crb3}
\end{equation}%
So, after inserting to Eq. (\ref{cra3}) and using the assumption (\ref{crg3}%
) one gets%
\begin{equation}
\left\langle {\bf nj}\right\rangle =\frac{1}{\Omega }\int \int_{\Omega }%
\left[ \left( G_{+1/2}(p_{0})-G_{-1/2}(p_{0})\right) /2+G(p_{0})\varepsilon
_{klm}n_{k}p_{l}x_{m}\right] d^{3}xd^{3}p  \label{crc3}
\end{equation}%
\newline
\[
=\frac{1}{2}\int \Delta G(p_{0})d^{3}p=\frac{1}{2}, 
\]%
since the term $\varepsilon _{klm}n_{k}p_{l}x_{m}$, due to spheric symmetry,
vanishes. One can check, if ${\bf n}^{\prime },{\bf n}^{\prime \prime }$ are
vectors, which together with ${\bf n}$ generate an orthonormal base in the
frame of centre of mass of the three fermions, then a similar calculation
gives%
\begin{equation}
\left\langle {\bf n}^{\prime }{\bf j}\right\rangle =\left\langle {\bf n}%
^{\prime \prime }{\bf j}\right\rangle =0.  \label{cx1}
\end{equation}%
Obviously, the simplest way is to use the base like:%
\begin{equation}
{\bf n}=(0,0,1),\qquad {\bf n}^{\prime }=(0,1,0),\qquad {\bf n}^{\prime
\prime }=(1,0,0).  \label{cx2}
\end{equation}%
Since we work with the probabilistic description (in terms of quantum
mechanics with the statistical mixture of states) by means of the
distributions $G_{\lambda }$, as a result we can obtain only the mean values
of the total spin projections $\left\langle {\bf J}\right\rangle =(0,0,1/2)$%
. Nevertheless one could consider a more rigorous (but more complicated)
approach, in which the three fermion system is not constructed as the
statistical mixture of plane waves, but as the composition of the three pure
states $j=1/2,$ $j_{z}=\pm 1/2$ with the condition, that the whole system
represents a pure state $J=1/2,$ $J_{z}=1/2$. These states are represented
by the relativistic spheric waves (spinors), which imply the corresponding
probabilistic distributions $G,G_{\lambda },\Delta G$ and $H$ have spheric
symmetry. In other words, if in our approach we assume the system in a pure
state $J=1/2$, then its probabilistic description in terms of the plane
waves will be defined by the distributions $G_{\lambda }$, which are
spherically symmetric. In fact, that is the reason, why we require spheric
symmetry, deformed distributions $G_{\lambda }$ would contradict the
eigenstate $J=1/2$.

Let us point out, in the relativistic case, having one fermion state with
definite projection ${\bf nj}$ of the total angular momentum, one cannot
separate its orbital and spin part (with exception of the special case when $%
{\bf n}\parallel \pm {\bf p}$), i.e. account with the fermion orbital
momentum is crucial for a consistent calculation of the resulting spin. On
the other hand, the similar calculation, in which the orbital part ${\bf l}$ 
$\ $is ignored, gives%
\[
\psi _{p,\lambda }^{\dagger }(x){\bf n\Sigma }\psi _{p,\lambda }(x)=\frac{1}{%
\Omega N}\left( \lambda \phi _{\lambda }^{\dagger }\phi _{\lambda }+\phi
_{\lambda }^{\dagger }\frac{{\bf p}{\bf \sigma }\cdot {\bf n\sigma }\cdot 
{\bf p}{\bf \sigma }}{2\left( p_{0}+m\right) ^{2}}\phi _{\lambda }\right) 
\]%
\[
=\frac{1}{\Omega N}\left( \lambda +\phi _{\lambda }^{\dagger }\frac{{\bf p}%
{\bf \sigma }\cdot \left( -{\bf p}{\bf \sigma }\cdot {\bf n\sigma }+2{\bf pn}%
\right) }{2\left( p_{0}+m\right) ^{2}}\phi _{\lambda }\right) 
\]%
\[
=\frac{1}{\Omega N}\left( \lambda -\lambda \frac{{\bf p}^{2}}{\left(
p_{0}+m\right) ^{2}}+\phi _{\lambda }^{\dagger }\frac{{\bf p}{\bf \sigma }%
\cdot {\bf pn}}{\left( p_{0}+m\right) ^{2}}\phi _{\lambda }\right) . 
\]%
Since%
\begin{equation}
{\bf p}{\bf \sigma }\cdot {\bf pn}=\sum_{i=1}^{3}p_{i}^{2}\sigma
_{i}n_{i}+\sum_{j\neq i}p_{i}p_{j}\sigma _{i}n_{j}  \label{ct1}
\end{equation}%
one can write%
\[
\left\langle {\bf n\Sigma }\right\rangle =\int \int_{\Omega }\sum_{\lambda
}G_{\lambda }(p_{0})\left( \psi _{p,\lambda }^{\dagger }(x){\bf n\Sigma }%
\psi _{p,\lambda }(x)\right) d^{3}xd^{3}p 
\]%
\[
=\int \sum_{\lambda }G_{\lambda }(p_{0})\frac{\lambda }{N}\left( 1-\frac{%
{\bf p}^{2}}{\left( p_{0}+m\right) ^{2}}+\frac{2{\bf p}^{2}}{3\left(
p_{0}+m\right) ^{2}}\right) d^{3}p, 
\]%
where inserting the formula (\ref{ct1}), we take into account, that due to
spheric symmetry the terms $p_{i}p_{j}$ ($j\neq i$) vanish and the terms $%
p_{i}^{2}$ can be substituted by ${\bf p}^{2}/3$. The last relation can be
further simplified: 
\begin{equation}
\left\langle {\bf n\Sigma }\right\rangle =\frac{1}{2}\int \Delta
G(p_{0})\left( \frac{1}{3}+\frac{2m}{3p_{0}}\right) d^{3}p\leq \frac{1}{2}.
\label{cs1}
\end{equation}%
One can observe, that the correspondence with Eq. (\ref{crc3}) takes place
only for the system of {\it static} fermions.

For further consideration, it will be useful to substitute the vector ${\bf n%
}$, representing the direction of the fermion polarization, by the
corresponding covariant polarization vector $w^{\sigma }(\lambda )$, which
satisfies%
\begin{equation}
w^{2}(\lambda )=-1,\qquad w(\lambda )\cdot p=0  \label{cs2}
\end{equation}%
and 
\begin{equation}
w(\lambda )=\frac{\lambda }{\left| \lambda \right| }(0,{\bf n});\qquad
\lambda =\pm \frac{1}{2}  \label{cx3}
\end{equation}%
in the fermion rest frame. The explicit representation of the vectors $%
w(\lambda )$ will be defined hereinafter.

Now, let us expose this system as a (fixed) target to the beam of polarized
electrons (e.g. $helicity$ $=+1/2$) coming with the momentum 
\begin{equation}
k=\left( k_{0},\sqrt{k_{0}^{2}-m_{e}^{2}},0,0\right)  \label{csa2}
\end{equation}%
and let us calculate the form of corresponding differential cross-section.
The spin dependent part of the cross-section for interaction with a single
fermion in one photon approximation has the form%
\begin{equation}
d\sigma \sim -L^{\alpha \beta (A)}(q,s)T_{\alpha \beta }^{(A)}.  \label{cr4}
\end{equation}%
The antisymmetric tensor $L^{\alpha \beta (A)}$, (see e.g. \cite{efrem1})
related to the electron beam reads:%
\begin{equation}
L^{\alpha \beta (A)}=m_{e}\varepsilon _{\alpha \beta \lambda \sigma
}s^{\lambda }q^{\sigma },  \label{cs3}
\end{equation}%
where $m_{e}$ is the electron mass, $s$ denotes its polarization vector%
\begin{equation}
s=\frac{1}{m_{e}}\left( \sqrt{k_{0}^{2}-m_{e}^{2}},k_{0},0,0\right) ;\qquad
s^{2}=-1,\qquad ks=0  \label{cs4}
\end{equation}%
and $\ q=k-k^{\prime }$ is the photon momentum. The antisymmetric tensor $%
T^{\alpha \beta (A)}$ related to the single fermion inside the target has a
similar form:%
\begin{equation}
T^{\alpha \beta (A)}=m\varepsilon _{\alpha \beta \lambda \sigma }q^{\lambda
}w^{\sigma }(\lambda ),  \label{cs5}
\end{equation}%
where $m$ and $w(\lambda )$ denote the fermion mass and polarization vector.
If one assumes, that the electron scattering can be described as the
incoherent sum of the interactions with the single plane waves, then the
tensor $T^{\alpha \beta (A)}$ reads%
\begin{equation}
T_{\alpha \beta }^{(A)}=\varepsilon _{\alpha \beta \lambda \sigma
}q^{\lambda }m\int H(p_{0})w^{\sigma }\delta ((p+q)^{2}-m^{2})\frac{d^{3}p}{%
p_{0}}.  \label{cr5}
\end{equation}%
Here the charge factors and the two possible signs of $w^{\sigma }$ are
included into the tensor through the distribution (\ref{cre3}). By the
symbol $w^{\sigma }$ we mean $w^{\sigma }(\lambda =+1/2)$. Let us remark,
this form of the antisymmetric part of the hadronic tensor is very similar
to that used in \cite{cfs}. Further, we can modify the $\delta -$function
term:%
\begin{equation}
\delta ((p+q)^{2}-m^{2})d^{3}p=\delta (2pq+q^{2})d^{3}p=\frac{1}{2\xi }%
\delta (\frac{pq}{\xi }+\frac{q^{2}}{2\xi })d^{3}p,  \label{cr9}
\end{equation}%
where $\xi $ is arbitrary constant, which only rescales the integration
variable. Now, let us imagine, that our target is a part of the greater
system, which is at rest with respect to the given reference frame and has
the mass $M$, but at the same time the probing electron interact only with
the three fermions. If we put%
\begin{equation}
\xi =Mq_{0}=M\nu ,  \label{cr10}
\end{equation}%
then in the $\delta -$function\ one can identify the terms known from the
formalism of deep inelastic scattering:%
\begin{equation}
-\frac{q^{2}}{2M\nu }=\frac{Q^{2}}{2M\nu }=x,  \label{cr11}
\end{equation}%
which is the Bjorken scaling variable, its value can be directly determined
using only initial and final momenta of the scattered electron. This
variable is in the $\delta -$function compensated by the ratio $pq/M\nu $,
which after boosting the whole target of mass $M$ to the infinite momentum
frame approximately represents ratio of dominating momenta components $%
p^{\prime }/P^{\prime }$ of the fermion and the target.

The explicit form of the polarization vector $w$ can be found as follows.
First, let us transform the vector $w=(0,{\bf n})$ from the fermion rest
frame to the target rest frame. After decomposition of the vector ${\bf n}$
to longitudinal and transversal parts with respect to the momentum fermion $%
{\bf p}$, the corresponding Lorentz boost gives%
\begin{equation}
(0,{\bf n})\rightarrow w=\left( \frac{{\bf pn}}{m},\,{\bf n}+\frac{{\bf pn}}{%
m(m+p_{0})}{\bf p}\right) .  \label{cr12}
\end{equation}%
Secondly, let us make a Lorentz boost of the whole target with mass $M$ to
some another frame, which is defined by the new components of the target
momentum 
\begin{equation}
\left( M,0,0,0\right) \rightarrow P=\left( P_{0},{\bf P}\right) ;\qquad
P^{2}=M^{2}.  \label{cr13}
\end{equation}%
Next, if we define the covariant vector $S$ by its components in the target
rest frame as 
\begin{equation}
S=(0,{\bf n}),  \label{cr14}
\end{equation}%
then the polarization vector $w$ can be written in manifestly covariant form%
\begin{equation}
w^{\sigma }=AP^{\sigma }+BS^{\sigma }+Cp^{\sigma },  \label{cr15}
\end{equation}%
where $A,B,C$ are invariant functions (scalars) of the vectors $P,S,p$.
These three functions are fixed by two the conditions (\ref{cs2}) and by the
constraint (\ref{cr12}) valid in the target rest frame. A simple calculation
gives:%
\begin{equation}
A=-\frac{pS}{pP+mM},\qquad B=1,\qquad C=\frac{M}{m}A.  \label{cr16}
\end{equation}%
So, we have obtained explicit covariant form of the polarization vector $w$
entering the tensor (\ref{cr5}), which can be now in accordance with the
relations (\ref{cr9})-(\ref{cr11}) rewritten%
\begin{equation}
T_{\alpha \beta }^{(A)}=\varepsilon _{\alpha \beta \lambda \sigma
}q^{\lambda }\frac{m}{2Pq}\int H\left( \frac{pP}{M}\right) w^{\sigma }\delta
\left( \frac{pq}{Pq}-x\right) \frac{d^{3}p}{p_{0}},  \label{cr17}
\end{equation}%
where we use the invariant term $Pq$ instead of $M\nu $ and $H(pP/M)$
instead of $H(p_{0})$.

On the other hand, in accordance with the general rule (see e.g. \cite%
{efrem1}), the antisymmetric tensor $T_{\alpha \beta }^{(A)}$ appearing in
the formula for the cross-section (\ref{cr4}), has the form%
\begin{equation}
T_{\alpha \beta }^{(A)}=\varepsilon _{\alpha \beta \lambda \sigma
}q^{\lambda }\left\{ MS^{\sigma }G_{1}+[(Pq)S^{\sigma }-(qS)P^{\sigma }]%
\frac{G_{2}}{M}\right\} ,  \label{cr18}
\end{equation}%
where $M,P,S$ represent the target mass, momentum and spin polarization
vector, which satisfies%
\begin{equation}
S^{2}=-1,\qquad PS=0.  \label{cra18}
\end{equation}%
The invariants $G_{1}$ and $G_{2}$ are the spin structure functions. In the
next we shall identify the parameters $M,P,S$ in Eq. (\ref{cr18}) with those
in the model described above and simultaneously we shall attempt to
determine the spin structure functions corresponding to our target. First of
all, we modify the Eq. (\ref{cr18}) by the substitution%
\begin{equation}
G_{S}=MG_{1}+\frac{Pq}{M}G_{2},\quad G_{P}=\frac{qS}{M}G_{2},  \label{cr19}
\end{equation}%
which gives%
\begin{equation}
T_{\alpha \beta }^{(A)}=\varepsilon _{\alpha \beta \lambda \sigma
}q^{\lambda }\left\{ S^{\sigma }G_{S}-P^{\sigma }G_{P}\right\} .
\label{cr20}
\end{equation}%
Comparison with Eq. (\ref{cr17}) gives the equation for the structure
functions:%
\begin{equation}
\varepsilon _{\alpha \beta \lambda \sigma }q^{\lambda }\left\{ S^{\sigma
}G_{S}-P^{\sigma }G_{P}\right\} =\varepsilon _{\alpha \beta \lambda \sigma
}q^{\lambda }\frac{m}{2Pq}\int H\left( \frac{pP}{M}\right) w^{\sigma }\delta
\left( \frac{pq}{Pq}-x\right) \frac{d^{3}p}{p_{0}}.  \label{cr21}
\end{equation}%
Because of the antisymmetry of the tensor $\varepsilon $ and after inserting
from the relation (\ref{cr15}) it follows that 
\begin{equation}
S^{\sigma }G_{S}-P^{\sigma }G_{P}=\frac{m}{2Pq}\int H\left( \frac{pP}{M}%
\right) \left( AP^{\sigma }+BS^{\sigma }+Cp^{\sigma }\right) \delta \left( 
\frac{pq}{Pq}-x\right) \frac{d^{3}p}{p_{0}}+Dq^{\sigma },  \label{cr25}
\end{equation}%
where $D$ is some scalar function and the functions $A,B,C$ are given by the
relations (\ref{cr16}). After contracting with $P_{\sigma },S_{\sigma }$ and 
$q_{\sigma }$ one gets the equations for unknown functions $G_{S},G_{P}$ and 
$D$:%
\begin{equation}
-M^{2}G_{P}=\frac{m}{2Pq}\int H\left( \frac{pP}{M}\right) \left(
AM^{2}+C\cdot pP\right) \delta \left( \frac{pq}{Pq}-x\right) \frac{d^{3}p}{%
p_{0}}+D\cdot Pq,  \label{cr26}
\end{equation}%
\begin{equation}
-G_{S}=\frac{m}{2Pq}\int H\left( \frac{pP}{M}\right) \left( -B+C\cdot
pS\right) \delta \left( \frac{pq}{Pq}-x\right) \frac{d^{3}p}{p_{0}}+D\cdot
qS,  \label{cr27}
\end{equation}%
\begin{equation}
qS\cdot G_{S}-Pq\cdot G_{P}=\frac{m}{2Pq}\int H\left( \frac{pP}{M}\right)
\left( A\cdot Pq+B\cdot qS+C\cdot pq\right)  \label{cr28}
\end{equation}%
\[
\times \delta \left( \frac{pq}{Pq}-x\right) \frac{d^{3}p}{p_{0}}+Dq^{2} 
\]%
and inserting $G_{P},G_{S}$ from the first two equations to the last one
gives the condition for $D$:%
\begin{equation}
\frac{m}{2Pq}\int H\left( \frac{pP}{M}\right) \left( C\cdot pu\right) \delta
\left( \frac{pq}{Pq}-x\right) \frac{d^{3}p}{p_{0}}+D\cdot qu=0,  \label{cr29}
\end{equation}%
where we denote%
\[
u\equiv q+\left( qS\right) S-\frac{\left( Pq\right) }{M^{2}}P. 
\]%
Finally, inserting $D$ from this equation to Eqs. (\ref{cr26}), (\ref{cr27})
gives with the use of relations (\ref{cr16}) the structure functions%
\begin{equation}
G_{P}=\frac{m}{2Pq}\int H\left( \frac{pP}{M}\right) \frac{pS}{pP+mM}\left[ 1+%
\frac{1}{mM}\left( pP-\frac{pu}{qu}Pq\right) \right] \delta \left( \frac{pq}{%
Pq}-x\right) \frac{d^{3}p}{p_{0}},  \label{cr30}
\end{equation}%
\begin{equation}
G_{S}=\frac{m}{2Pq}\int H\left( \frac{pP}{M}\right) \left[ 1+\frac{pS}{pP+mM}%
\frac{M}{m}\left( pS-\frac{pu}{qu}qS\right) \right] \delta \left( \frac{pq}{%
Pq}-x\right) \frac{d^{3}p}{p_{0}}.  \label{cr31}
\end{equation}%
The spin structure functions in the standard notation $g_{1}=M\cdot Pq\cdot
G_{1},$ $g_{2}=\left( Pq\right) ^{2}/M\cdot G_{2}$ can be now obtained from
Eqs. (\ref{cr19}):%
\begin{equation}
g_{1}=Pq\left( G_{S}-\frac{Pq}{qS}G_{P}\right) ,\qquad g_{2}=\frac{\left(
Pq\right) ^{2}}{qS}G_{P},\qquad g_{1}+g_{2}=PqG_{S},  \label{cra31}
\end{equation}%
where the functions $G_{S},G_{P}$ are given by relations (\ref{cr30}), (\ref%
{cr31}). Corresponding integrals, as shown in the Appendix, can be
simplified to the form (\ref{c9}), (\ref{c10}). Let us remark, resulting
functions $g_{1},g_{2}$, after inserting from the relations (\ref{c9}), (\ref%
{c10}) into Eq. (\ref{cra31}) do not depend on the variable $qS$ despite the
fact, that such terms are present in the starting integrals (\ref{cr30}), (%
\ref{cr31}) in a non-trivial way. This is a consequence of spheric symmetry
of the distribution $H$, which as we have suggested, follows from the
requirement $J=1/2$.

\section{Sum rules}

\label{sec3} For next analysis of the obtained structure functions it is
convenient to express the integrals (\ref{cr30}),(\ref{cr31}) in the target
rest frame, where $P=(M,0,0,0)$ and $S=(0,{\bf n})$. Detailed calculation is
done in the Appendix. Now, let us assume $Q^{2}\gg 4M^{2}x^{2}$, then%
\[
\frac{\left| {\bf q}\right| }{\nu }=\sqrt{1+4M^{2}x^{2}/Q^{2}}\rightarrow 1 
\]%
and using the second relation (\ref{cra31}) and Eq. (\ref{cb4}) one gets%
\begin{equation}
\Gamma _{2}\equiv \int g_{2}(x)dx=-\pi \int \int H(p_{0})\left( p_{1}+\frac{%
p_{1}^{2}-p_{T}^{2}/2}{p_{0}+m}\right) \delta \left( \frac{p_{0}+p_{1}}{M}%
-x\right) \frac{p_{T}dp_{1}dp_{T}}{p_{0}}dx  \label{sr1}
\end{equation}%
\[
=-\pi \int H(p_{0})\left( p_{1}+\frac{p_{1}^{2}-p_{T}^{2}/2}{p_{0}+m}\right) 
\frac{p_{T}dp_{1}dp_{T}}{p_{0}}. 
\]%
In the last integral, due to spheric symmetry of the distribution $H$, the
terms proportional to $p_{1}$ and $p_{1}^{2}-p_{T}^{2}/2$ vanish, insofar
that%
\begin{equation}
\Gamma _{2}=0,  \label{sr2}
\end{equation}%
which is the known Burkhardt-Cottingham sum rule \cite{buco}. Similarly the
third relation (\ref{cra31}) and Eq. (\ref{cc4}) give%
\begin{equation}
\int \left( g_{1}(x)+g_{2}(x)\right) dx=\pi \int H(p_{0})\left( m+\frac{%
p_{T}^{2}}{2\left( p_{0}+m\right) }\right) \frac{p_{T}dp_{1}dp_{T}}{p_{0}}.
\label{sr4}
\end{equation}%
After the substitution%
\begin{equation}
2\pi p_{T}dp_{1}dp_{T}=d^{3}p  \label{sra4}
\end{equation}%
and using relation (\ref{sr2}) one gets%
\begin{equation}
\Gamma _{1}\equiv \int g_{1}(x)dx=\frac{1}{2}\int H(p_{0})\left( m+\frac{%
{\bf p}^{2}}{3\left( p_{0}+m\right) }\right) \frac{d^{3}p}{p_{0}}.
\label{sr5}
\end{equation}%
Simple modification then gives%
\begin{equation}
\Gamma _{1}=\frac{1}{2}\int H(p_{0})\left( \frac{1}{3}+\frac{2m}{3p_{0}}%
\right) d^{3}p.  \label{sr6}
\end{equation}%
More detailed analysis of this result will be done in the next section.

The relations (\ref{cb4}) and (\ref{cc4}) \ can be used also for the
calculation of the higher momenta. Generally, if $F$ is a function defined as%
\[
F(x)=\int K(p)\delta \left( \frac{p_{0}+p_{1}}{M}-x\right) d^{3}p, 
\]%
then%
\[
\int x^{n}F(x)dx=\int \int K(p)x^{n}\delta \left( \frac{p_{0}+p_{1}}{M}%
-x\right) d^{3}pdx 
\]%
\[
=\int \int K(p)\left( \frac{p_{0}+p_{1}}{M}\right) ^{n}\delta \left( \frac{%
p_{0}+p_{1}}{M}-x\right) d^{3}pdx 
\]%
\[
=\int K(p)\left( \frac{p_{0}+p_{1}}{M}\right) ^{n}d^{3}p. 
\]%
Application of this rule to Eqs. (\ref{cb4}) and (\ref{cc4}) gives after the
substitution (\ref{sra4}) and with the use of the second and third relation (%
\ref{cra31}):%
\begin{equation}
\int xg_{2}dx=-\frac{1}{6M}\int H(p_{0})\left( p_{0}-\frac{m^{2}}{p_{0}}%
\right) d^{3}p,  \label{sr7}
\end{equation}%
\begin{equation}
\int x\left( g_{1}+g_{2}\right) dx=\frac{1}{6M}\int H(p_{0})\left(
p_{0}+2m\right) d^{3}p.  \label{sr8}
\end{equation}%
These equalities imply relation%
\begin{equation}
\int x\left( g_{1}+2g_{2}\right) dx=\frac{1}{6M}\int H(p_{0})\left( 2m+\frac{%
m^{2}}{p_{0}}\right) d^{3}p,  \label{sr9}
\end{equation}%
which in the limit of the massless fermions coincides with the Efremov -
Leader - Teryaev (ELT) sum rule \cite{efrem2}:%
\begin{equation}
\int x\left( g_{1}+2g_{2}\right) dx=0.  \label{sr10}
\end{equation}

\section{Discussion}

\label{sec4} In the previous sections we have studied the properties of the
spin structure functions related to the system of quasifree fermions on mass
shell. This system can be compared with the na\"{\i}ve QPM, which is with
embedded QCD corrections\ yet the basic tool for the analysis and
interpretation of polarized and unpolarized deep inelastic scattering data.
What is the difference between our approach and the na\"{\i}ve QPM,\ if one
speaks about the proton spin structure functions? To simplify this
discussion, let us assume:

1) Spin contribution from the sea of quark-antiquark pairs and gluons can be
neglected. Then the three fermions in our approach correspond to the three
proton valence quarks. So, in this simplified scenario, the proton spin is
generated only by the valence quarks.

2) In an accordance with the non-relativistic {\it SU(6)} approach the spin
contribution of individual valence terms is given as%
\begin{equation}
s_{u}=4/3,\qquad s_{d}=-1/3.  \label{d1}
\end{equation}%
Let us point out, in the given context the term valence quarks means nothing
else, than the three fermions with defined momenta distribution, charge,
mass and polarization.

Then according to the na\"{\i}ve {\it SU(6)} version of the QPM we have%
\begin{equation}
g_{1}(x)=\frac{1}{2}\sum e_{j}^{2}\Delta q_{j}(x)=\frac{1}{2}\left( \left( 
\frac{2}{3}\right) ^{2}\frac{2}{3}u_{val}(x)-\left( \frac{1}{3}\right) ^{2}%
\frac{1}{3}d_{val}(x)\right) ,  \label{d2}
\end{equation}%
corresponding to two the quarks with distribution $u_{val}(x)$ and the one
with distribution $d_{val}(x)$, which are normalized as%
\begin{equation}
\frac{1}{2}\int u_{val}(x)dx=\int d_{val}(x)dx=1.  \label{d3}
\end{equation}%
It follows, that%
\begin{equation}
\Gamma _{1}=\int g_{1}(x)dx=\frac{5}{18}\doteq 0.28.  \label{d4}
\end{equation}%
This number overestimates more than twice the experimental value.
Disagreement is generally interpreted as a contradiction with the
assumption, that the proton spin is generated only by spins of the valence
quarks.

Now let us calculate the $\Gamma _{1}$ in our approach. Let us denote
momenta distributions of the valence quarks in the target rest frame by
symbols $h_{u}$ and $h_{d}$ with the normalization%
\begin{equation}
\frac{1}{2}\int h_{u}(p_{0})d^{3}p=\int h_{d}(p_{0})d^{3}p=1.  \label{d5}
\end{equation}%
These distributions are connected with the $u_{val}(x)$ and $d_{val}(x)$
defined above by the relation%
\begin{equation}
q_{val}(x)=\int h_{q}(p_{0})\delta \left( \frac{p_{0}\nu +p_{1}\left| {\bf q}%
\right| }{M\nu }-x\right) d^{3}p.  \label{d6}
\end{equation}%
The charge weighted distribution (\ref{cre3}), in an {\it SU(6)} picture,
reads%
\begin{equation}
H(p_{0})=\sum e_{j}^{2}\Delta h_{j}(p_{0})=\left( \left( \frac{2}{3}\right)
^{2}\frac{2}{3}h_{u}(p_{0})-\left( \frac{1}{3}\right) ^{2}\frac{1}{3}%
h_{d}(p_{0})\right) .  \label{d7}
\end{equation}%
Now, for simplicity let us assume the same shape of the distributions for
both the flavours: 
\begin{equation}
\frac{1}{2}h_{u}(p_{0})=h_{d}(p_{0})\equiv h(p_{0}).  \label{da8}
\end{equation}%
Then it follows%
\begin{equation}
G(p_{0})=3h(p_{0}),\qquad \Delta G(p_{0})=h(p_{0}),\qquad H(p_{0})=\frac{5}{9%
}h(p_{0})  \label{db8}
\end{equation}%
and the relations (\ref{cs1}) and (\ref{sr6}) can be rewritten:%
\begin{equation}
\left\langle {\bf n\Sigma }\right\rangle =\frac{1}{2}\int h(p_{0})\left( 
\frac{1}{3}+\frac{2m}{3p_{0}}\right) d^{3}p,  \label{dc8}
\end{equation}%
\begin{equation}
\Gamma _{1}=\frac{5}{18}\int h(p_{0})\left( \frac{1}{3}+\frac{2m}{3p_{0}}%
\right) d^{3}p.  \label{dd8}
\end{equation}%
These relations imply:

{\it a)} Because the distribution $h$ has the defined normalization, the
corresponding integrals reach their maximum in the limit, when the fermions
are static ($p_{0}=m$). On the other hand in the limit of massless fermions (%
$m\ll p_{0}$) these integrals represent only one third of their maximal
value. In particular, the $\Gamma _{1}$ satisfies: 
\begin{equation}
\frac{5}{18}\geq \Gamma _{1}\geq \frac{5}{54}.  \label{d10}
\end{equation}

{\it b)} Both the integrals are (up to the factor $5/9$) equal. It follows,
that in the case of non static fermions the $\Gamma _{1}$ ''measures'' only
the contribution from their spins, which is only part of the their angular
momenta, see derivation of the relation (\ref{cs1}). Fermions with momentum $%
{\bf p\neq 0}$, which is not parallel to $\pm {\bf n}$, necessarily
contribute to the total angular momentum also by some orbital part.

Further let us notice, if we denote%
\begin{equation}
\gamma _{ELT}\equiv \int x\left( g_{1}+2g_{2}\right) dx,  \label{da10}
\end{equation}%
then Eq. (\ref{sr9}) and the third relation (\ref{db8}) imply%
\begin{equation}
\frac{2}{3}m\leq \frac{18}{5}\gamma _{ELT}\cdot M\leq m.  \label{db10}
\end{equation}

Why these two very simple approaches for description of the target
consisting of the three fermions differ so strongly regarding the prediction 
$\Gamma _{1}$? The reason is following. The standard formulation of the QPM
is closely connected with the preferred reference system - infinite momentum
frame (IMF). The basic relations between the distribution and structure
functions like%
\begin{equation}
g_{1}(x)=\frac{1}{2}\sum e_{j}^{2}\Delta q_{j}(x),\qquad F_{2}(x)=x\sum
e_{i}^{2}q_{i}(x)  \label{d11}
\end{equation}%
are derived with the use of approximation%
\begin{equation}
p_{\alpha }=xP_{\alpha },  \label{d12}
\end{equation}%
which seems to be plausible in the IMF. Nevertheless, in the covariant
formulation this relation is equivalent to the assumption, that the quarks
are static with respect to the proton, since the velocities $p_{j}/p_{0}$
and $P_{j}/P_{0}$ are the same. In the proton rest frame it means ${\bf p}=%
{\bf 0}$. That is why both the approaches are equivalent for the static
quarks but differ for the quarks, which have some intrinsic motion inside
the proton. In our approach we do not use assumption (\ref{d12}) and as a
result if $p_{\alpha }\neq xP_{\alpha }$ we obtain different relations
between the distribution and structure functions. In other words, the fact,
that the experimental value $\Gamma _{1}$ is substantially under the value
predicted by the na\"{\i}ve QPM in standard formulation, can be in our
approach interpreted as a direct consequence of the quark intrinsic motion.

Of course, the approach discussed above concerns the simplified scenario of
the quasifree fermions on mass shell. Na\"{\i}ve QPM\ represents only a
first approximation for a description of real nucleon, but the consistent
accounting for the quark intrinsic motion as suggested in our approach can,
in some aspects, improve this approximation considerably.

Nevertheless, in the realistic case of partons inside the nucleon the
situation is still much more delicate. The interaction among the quarks and
gluons is very strong, partons themselves are mostly in some shortly living
virtual states, is it possible to speak about their mass at all? Strictly
speaking probably not. The mass in the exact sense is well defined only for
free particles, whereas the partons are never free. However one can assume
the following. The relations obtained in the previous sections can be used
as a good approximation even for the interacting quarks, but provided that
the term {\it mass of quasifree parton} is substituted by the term {\it %
parton effective mass}. By this term we mean the mass, which a free parton
would have to have to interact with the probing photon equally as the real,
bounded one. Intuitively, this mass should correlate to $Q^{2}$: a lower $%
Q^{2}$ roughly means, that the photon ''sees'' the quark surrounded by some
cloud of gluons and quark-antiquark pairs as a one particle - by which this
photon is absorbed. And on contrary, the higher $Q^{2}$ should mediate
interaction with more ''isolated'' quark. Moreover, one should accept that
the value of the effective mass can even for a fixed $Q^{2}$ fluctuate. Such
phenomenological model was suggested in \cite{zav3}, but unfortunately
calculation was based on the form of quark polarization vector which is not
correct. Despite of that, the general considerations in mentioned paper can
be sensible. Corresponding numeric recalculation with the correct input
obtained in the present study for the invariants $A,B,C,D$ [relations (\ref%
{cr16}),(\ref{cr29})] should be done in a separate paper.

\section{Summary and conclusion}

In the present paper we have studied the spin structure functions of the
system of quasifree fermions on mass shell and with spherically symmetric
distribution of their momenta. The main results can be summarized as follows:

1) Using consistently covariant description of this simple system, we have
shown how the structure functions depend on the intrinsic motion of the
fermions. In particular, we have suggested, that the momenta $\Gamma _{1}$
corresponding to the two extreme scenarios, of the static (massive) fermions
and massless fermions, can differ significantly: $\Gamma _{1}(m\ll
p_{0})/\Gamma _{1}(p_{0}\approx m)=1/3$.

2) We have shown, what sum rules follow from the obtained spin structure
functions. Further we have shown, how these rules are related to some sum
rules well known from the QPM phenomenology.

3) We have done a comparison with the corresponding relations for the
structure functions following from the standard formulation of the na\"{\i}%
ve QPM. Both the approaches are basically equivalent for the static quarks.
Differences for quarks with intrinsic motion inside the proton are result of
the conflict with the assumption $p_{\alpha }=xP_{\alpha }$, which is
crucial for derivation of the relations between structure and distribution
functions in the standard QPM.

4) The difference between the experimental value $\Gamma _{1}$ for the
proton and the corresponding value expected from the na\"{\i}ve QPM, or at
least a part of this difference, can be interpreted as a consequence of the
quark motion inside the proton. \\[5mm]
\noindent {\bf Acknowledgement:} I would like to thank Anatoli Efremov and
Oleg Teryaev for many useful discussions and valuable comments.

\bigskip

\appendix

\section{Calculation of the integrals related to $G_{P},G_{S}$}

\label{appe2}Integrals in the relations (\ref{cr30}), (\ref{cr31}) expressed
in the target rest frame read%
\begin{equation}
G_{P}=-\frac{m}{2M^{2}\nu }\int H(p_{0})  \label{ca}
\end{equation}%
\[
\times \frac{{\bf pn}}{p_{0}+m}\left[ 1+\frac{1}{m}\left( p_{0}-\frac{{\bf p}%
{\bf q}-\left( {\bf pn}\right) \left| {\bf q}\right| \cos \omega }{{\bf q}%
^{2}\sin ^{2}\omega }\nu \right) \right] \delta \left( \frac{pq}{M\nu }%
-x\right) \frac{d^{3}p}{p_{0}}, 
\]%
\begin{equation}
G_{S}=\frac{m}{2M\nu }\int H(p_{0})  \label{cb}
\end{equation}%
\[
\times \left[ 1+\frac{{\bf pn}}{p_{0}+m}\frac{1}{m}\left( {\bf pn}-\frac{%
{\bf p}{\bf q}-\left( {\bf pn}\right) \left| {\bf q}\right| \cos \omega }{%
{\bf q}^{2}\sin ^{2}\omega }\left| {\bf q}\right| \cos \omega \right) \right]
\delta \left( \frac{pq}{M\nu }-x\right) \frac{d^{3}p}{p_{0}}, 
\]%
where $\cos \omega \equiv {\bf qn}/\left| {\bf q}\right| .$\ For integration
we use the orthonormal system in which 
\begin{equation}
{\bf p}=p_{1}{\bf e}_{1}+p_{2}{\bf e}_{2}+p_{3}{\bf e}_{3},\quad {\bf e}%
_{1}=-\frac{{\bf q}}{\left| {\bf q}\right| },\quad {\bf e}_{2}=\frac{{\bf n}%
-({\bf ne}_{1}){\bf e}_{1}}{\sqrt{1-({\bf ne}_{1})^{2}}},\quad {\bf e}_{3}=%
{\bf e}_{1}\times {\bf e}_{2},  \label{c1}
\end{equation}%
so one gets 
\begin{equation}
{\bf p}{\bf q}=-p_{1}\left| {\bf q}\right| ,\qquad {\bf pn}=-p_{1}\cos
\omega +p_{2}\sin \omega ,\qquad \cos \omega \equiv \frac{{\bf qn}}{\left| 
{\bf q}\right| }.  \label{c2}
\end{equation}%
After the substitution $p_{2}=p_{T}\cos \varphi ,\ p_{3}=p_{T}\sin \varphi $
and taking into account that the terms proportional to $\cos \varphi $
disappear, the integrals can be rewritten%
\begin{equation}
G_{P}=\frac{\cos \omega }{2M^{2}\nu }\int H(p_{0})\left( p_{1}+\frac{\nu }{%
\left| {\bf q}\right| }\frac{p_{1}^{2}-p_{T}^{2}\cos ^{2}\varphi }{p_{0}+m}%
\right)  \label{c3}
\end{equation}%
\[
\times \delta \left( \frac{p_{0}\nu +p_{1}\left| {\bf q}\right| }{M\nu }%
-x\right) \frac{p_{T}dp_{1}dp_{T}d\varphi }{p_{0}}, 
\]%
\begin{equation}
G_{S}=\frac{m}{2M\nu }\int H(p_{0})\left( 1+\frac{p_{T}^{2}\cos ^{2}\varphi 
}{m\left( p_{0}+m\right) }\right) \delta \left( \frac{p_{0}\nu +p_{1}\left| 
{\bf q}\right| }{M\nu }-x\right) \frac{p_{T}dp_{1}dp_{T}d\varphi }{p_{0}},
\label{c4}
\end{equation}%
where $p_{0}=\sqrt{m^{2}+p_{T}^{2}+p_{1}^{2}}$. Integration over $\varphi $
gives%
\begin{equation}
G_{P}=\frac{\pi \cos \omega }{M^{2}\nu }\int H(p_{0})\left( p_{1}+\frac{\nu 
}{\left| {\bf q}\right| }\frac{p_{1}^{2}-p_{T}^{2}/2}{p_{0}+m}\right) \delta
\left( \frac{p_{0}\nu +p_{1}\left| {\bf q}\right| }{M\nu }-x\right) \frac{%
p_{T}dp_{1}dp_{T}}{p_{0}},  \label{cb4}
\end{equation}%
\begin{equation}
G_{S}=\frac{\pi m}{M\nu }\int H(p_{0})\left( 1+\frac{p_{T}^{2}/2}{m\left(
p_{0}+m\right) }\right) \delta \left( \frac{p_{0}\nu +p_{1}\left| {\bf q}%
\right| }{M\nu }-x\right) \frac{p_{T}dp_{1}dp_{T}}{p_{0}}.  \label{cc4}
\end{equation}%
Further, using the relation 
\begin{equation}
\frac{\left| {\bf q}\right| }{\nu }=\sqrt{1+4M^{2}x^{2}/Q^{2}}  \label{ca4}
\end{equation}%
one can check, that the argument of $\delta -$ function equals zero for 
\begin{equation}
p_{1}=\tilde{p}_{1}\equiv \frac{Mx-m_{T}^{2}/Mx}{\sqrt{1+4m_{T}^{2}/Q^{2}}+%
\sqrt{1+4M^{2}x^{2}/Q^{2}}},\qquad m_{T}^{2}\equiv m^{2}+p_{T}^{2}.
\label{c5}
\end{equation}%
This is the first root of the corresponding quadratic equation, the second
one is excluded, since in the effect of the $\delta -$ function this root is
compatible only with negative energy $p_{0}$. The energy corresponding to
the root (\ref{c5}) is 
\begin{equation}
p_{0}=\tilde{p}_{0}\equiv Mx-\frac{\tilde{p}_{1}\left| {\bf q}\right| }{\nu }%
=Mx-\tilde{p}_{1}\sqrt{1+4M^{2}x^{2}/Q^{2}}.  \label{c6}
\end{equation}%
Then in an accordance with the rule 
\begin{equation}
\delta (f(x))dx=\sum_{j}\frac{\delta (x-x_{j})}{\left| f^{\prime
}(x_{j})\right| }dx,\qquad f(x_{j})=0  \label{c7}
\end{equation}%
the $\delta -$ function in the integrals can be rewritten 
\begin{equation}
\delta \left( \frac{p_{0}\nu +p_{1}\left| {\bf q}\right| }{M\nu }-x\right)
dp_{1}=\frac{M\delta (p_{1}-\tilde{p}_{1})dp_{1}}{\tilde{p}_{1}/\tilde{p}%
_{0}+\sqrt{1+4M^{2}x^{2}/Q^{2}}}  \label{c8}
\end{equation}%
and afterwards the integrals are simplified 
\begin{equation}
G_{P}=\frac{\pi \cos \omega }{M\nu }\int_{0}^{p_{T\max }}H(\tilde{p}%
_{0})\left( \tilde{p}_{1}+\frac{\nu }{\left| {\bf q}\right| }\frac{\tilde{p}%
_{1}^{2}-p_{T}^{2}/2}{\tilde{p}_{0}+m}\right) \frac{p_{T}dp_{T}}{\tilde{p}%
_{1}+\tilde{p}_{0}\sqrt{1+4M^{2}x^{2}/Q^{2}}},  \label{c9}
\end{equation}%
\begin{equation}
G_{S}=\frac{\pi m}{\nu }\int_{0}^{p_{T\max }}H(\tilde{p}_{0})\left( 1+\frac{%
p_{T}^{2}/2}{m\left( \tilde{p}_{0}+m\right) }\right) \frac{p_{T}dp_{T}}{%
\tilde{p}_{1}+\tilde{p}_{0}\sqrt{1+4M^{2}x^{2}/Q^{2}}},  \label{c10}
\end{equation}%
where $\tilde{p}_{1}$ and $\tilde{p}_{0}$ depend on $p_{T}$\ according to
Eqs. (\ref{c5}) and (\ref{c6}). For the numeric calculation one should know
the upper limit $p_{T\max }$ for given $x,Q^{2}$ and $\tilde{p}_{0\max }$.
After inserting $\tilde{p}_{1}$ from Eq. (\ref{c5}) into Eq. (\ref{c6}) one
gets equation for $m_{T}^{2}$ 
\begin{equation}
\frac{\tilde{p}_{0\max }-Mx}{\sqrt{1+4M^{2}x^{2}/Q^{2}}}=-\frac{%
Mx-m_{T}^{2}/Mx}{\sqrt{1+4m_{T}^{2}/Q^{2}}+\sqrt{1+4M^{2}x^{2}/Q^{2}}}.
\label{c11}
\end{equation}%
Instead of $m_{T}^{2}$ it is useful to solve this equation first for $y=%
\sqrt{1+4m_{T}^{2}/Q^{2}\text{ }}$ obtaining the two roots 
\begin{equation}
y_{\pm }=\frac{A\pm \sqrt{A^{2}+4a(\tilde{p}_{0\max }+a)}}{2a},\qquad
A\equiv \frac{\tilde{p}_{0\max }-Mx}{\sqrt{1+4M^{2}x^{2}/Q^{2}}},\qquad
a\equiv \frac{Q^{2}}{4Mx}.  \label{c12}
\end{equation}%
Since $y_{-}<0,$ this root is excluded. The second root $y_{+\text{ }}$
after some computation implies 
\begin{equation}
m_{T\max }^{2}=Mx(2\tilde{p}_{0\max }-Mx)+\frac{(\tilde{p}_{0\max }-Mx)^{2}}{%
1+Q^{2}/4M^{2}x^{2}},\qquad p_{T\max }=\sqrt{m_{T\max }^{2}-m^{2}}.
\label{c13}
\end{equation}%
In this way we have the recipe how to calculate the integrals related to the
structure functions $G_{P},G_{S}$ corresponding to the distribution $%
H(p_{0})d^{3}p$.

\end{document}